\documentstyle[12pt,epsfig,color]{article}
\newcommand{\spacing}[1]{\renewcommand{\baselinestretch}{#1}\large\normalsize}
\newcommand{\dd}{{\rm d}}
\spacing{1.5}
\begin{document}

\begin{center}
{\large\bf{Implications of correlations and fluctuations in small systems}} \\
\bigskip

{\small
Debasish Das$^{a}$\footnote{email : debasish.das@saha.ac.in,dev.deba@gmail.com}, 
\medskip

$^a$Saha Institute of Nuclear Physics, A CI of Homi Bhabha National Institute,\\
 1/AF, Bidhan Nagar, Kolkata 700064, INDIA.\\
}

\end{center}
\date{\today}
\begin{abstract}

The small collision systems like p+p and p+A collisions have shown new features like
A+A collisions in the relativistic regime. These new aspects in small systems which
have altered our research and understanding on the two-particle
correlation measurements have been provided. Additionally, a critical observation of
the fluctuation measurements provides new ways to infer such novel happening in the
small collision systems. The ongoing and future endeavors towards the new
measurements are also discussed.

\end{abstract}


\section{\bf{Introduction}}

Quantum Chromodynamics (QCD), which is the standard theory of 
strong interactions, can help in understanding most of the 
visible matter of the universe~\cite{Deur:2016tte,Marciano:1977su,Marciano:1979wa,Kisslinger:2014uda}.
The QCD phase diagram has close relation to the history of the universe 
and can be probed by heavy-ion collisions. At very high temperature and density, 
the colour force between quarks~\cite{Riordan:1992hr} becomes so small that they can start behaving 
like essentially non-interacting free particles. Under such extreme conditions~\cite{Berges:1998nf,Kogut:2002kk,Fukushima:2011jc,Wilczek:1999ym},
the quarks tend to lose their confinement and a phase transition to a new form of matter, known as the 
Quark-Gluon Plasma (QGP) is formed~\cite{Kisslinger:2014uda}. It is widely believed that with the help of 
heavy-ion colliders like Relativistic Heavy Ion Collider (RHIC) and 
Large Hadron Collider (LHC),  it is possible to achieve this de-confined state~\cite{Kisslinger:2014uda,Fukushima:2011jc,Wilczek:1999ym}. 
The critical issues of de-confinement and the chiral transitions~\cite{Satz:1997bs}
which determine the QCD phase diagram can be studied in 
heavy-ion collision experiments~\cite{Kisslinger:2014uda,Fukushima:2011jc,Wilczek:1999ym,Rischke:2003mt,Braun-Munzinger:2015hba}. 
Among the other main challenges in this research domain is to simultaneously determine 
the initial temperature and the energy density of the matter created in such a collision
and hence the number of thermodynamic degrees of freedom~\cite{Kisslinger:2014uda,Cho:2008wg}.

Ultra-relativistic  collisions  can produce a small
droplet of de-confined quark matter~\cite{Shuryak:1980tp} in the form of strongly coupled QGP (sQGP), which allows 
the study of QCD in its non-perturbative regime~\cite{Heinz:2008tv}.  The energetic quarks
and gluons (hard partons)~\cite{Riordan:1992hr,Ellis:2014rma,Richard:2012xw} are 
created by perturbative processes at the same time~\cite{Riordan:1992hr,Ellis:2014rma,Richard:2012xw,Bruno:2014rxa}.
Since the QGP created in the relativistic heavy-ion collisions is not a weakly coupled gas but more a strongly
coupled liquid~\cite{Shuryak:2003xe,Liao:2006ry,Gyulassy:2004zy,Shuryak:2008eq,Nagle:2006cj}, 
the term sQGP was felt to be more apt~\cite{Heinz:2008tv,Shuryak:2005pp}.
After a few microseconds the universe seem to leave the QGP phase, with the available 
quarks and gluons combining towards the formation of the mesons and baryons~\cite{Heinz:2008tv}.
Understanding the early universe and its dynamics in terms of the Big Bang can be studied
experimentally by relativistic nucleus-nucleus collisions at such accelerator facilities, 
in terms of ``little bangs'' in the laboratory~\cite{Heinz:2008tv,Ellis:2005xq,Busza:2018rrf}.
The structure and evolution of the matter created in ``little bangs'' need to be closely 
characterized, so that we can explore the phase structure of the strongly
interacting matter~\cite{Heinz:2008tv,Ellis:2005xq,Busza:2018rrf,Bass:1998vz,Nagle:2018nvi}.

Hubble's discovery that the far away galaxies are moving away from us, with redshifts proportional to 
their relative distances from us, was the first to establish the theory of the expanding universe~\cite{Riess:2019qba,Allahverdi:2020bys}. 
There is necessary  evidence that the Universe at the initial epoch, was 3000 times smaller and hotter than today, 
if we extrapolate the present expansion backwards as explained by the Cosmic Microwave Background (CMB) Radiation~\cite{Gawiser:2000az,Durrer:2015lza}. 
In comparison to the present times, the state of the early universe was very different. 
Formation of the hadrons from the quarks and gluons was not possible 
and was seemingly filled with a thermalized plasma which is composed of de-confined quarks, anti-quarks, 
gluons and leptons and known as the primordial QGP~\cite{Ellis:2005xq}.  The initial conditions for the hot Big Bang 
are felt to have been produced by the quantum fluctuations during a period of inflationary expansion~\cite{Olive:2002qg,Linder:2008pp,Brandenberger:2003vk}.
In the early universe the matter was almost unequivocally QGP until the temperature reduced below a 
few trillion degrees, a millionth of a second after the Big Bang~\cite{Heinz:2008tv,Ellis:2005xq,Busza:2018rrf}. 
The understanding of correlations and fluctuations which are commonly applied for characterizing 
a physical system and providing information about the effective degrees of freedom, hence became important for 
deciphering varied aspects of early universe~\cite{Weinberg:2006ac,Baumann:2009ds}.
A notable example where the measurement of correlations has lead to scientific advancement,
are the fluctuations of the cosmic microwave background radiation~\cite{Baumann:2009ds}, 
first measured by the COBE satellite~\cite{Boggess:1992xla,Bennett:1996ce} and later through refined studies 
by WMAP~\cite{WMAP:2003elm,WMAP:2003ivt} and Planck collaborations~\cite{Planck:2015fie,Planck:2013pxb}. 
In the case of the CMB, the fluctuations are on the 
order of $10^{-4}$ with respect to the thermal distribution~\cite{Allahverdi:2020bys,Weinberg:2006ac,Baumann:2009ds}. 
Also, in the CMB one observes spatial correlations~\cite{Baumann:2009ds,Aurich:2014via}.

 However, while comparing with the case in heavy-ion collisions where we study QGP, we are restricted to 
correlations in momenta and also the association to spatial correlations is also non-trivial~\cite{Koch:2008ia}. 
Since fluctuations also assume paramount importance for studying perturbation of a thermodynamic system, some 
similarities also exist between the characterization methodologies, regarding the early universe and the heavy-ion collisions~\cite{Csernai:2012zf,Karsch:2005ps,Koch:2001zn,Qiu:2011iv,Lacey:2005qq,Becattini:2021lfq} .
A QGP system~\cite{Adams:2005dq,Adcox:2004mh,Back:2004je,Arsene:2004fa,Abelev:2014ffa,Krintiras:2020imy,Armesto:2015ioy} 
in a small span of time cools-down and eventually re-hadronizes. 
The study of the QGP typically involves analyses of soft (low $p_{T}$ ) 
particles whose behavior can be described by hydrodynamical models~\cite{Heinz:2008tv}, as well as those hard
partons that can interact with the QGP~\cite{Matsui:1986dk,Satz:1983jp,He:2014cla,Foka:2016zdb}. 
The ability to bridge these two regimes offers promise 
for understanding the interactions between the hard and soft scales\cite{Satz:1997bs}. 
As the de-confined phase is very short lived, direct detection is impossible.
Henceforth, the measurement of a quantity which is specific for the QGP phase, 
is required to conclude whether the phase transition occurred.~\cite{Kisslinger:2014uda,Rischke:2003mt,Braun-Munzinger:2015hba}. 
When the system undergoes a phase transition, a number of thermodynamic 
quantities show varying fluctuation patters, since fluctuations are sensitive to the nature of the transition~\cite{Koch:2008ia,Csernai:2012zf,Karsch:2005ps,Koch:2001zn}.
The QGP also shows collective behavior~\cite{Nagle:2018nvi,Voloshin:2008dg,Snellings:2011sz,Foka:2016vta}
that arises from the many-body interactions in QCD.  
To investigate the first order we observe and infer upon the thermal spectra and radial flow~\cite{Foka:2016zdb,Voloshin:2008dg,Snellings:2011sz}. 
Furthermore, a large quadrupole correlation due to elliptic flow ($v_2$) is also perceived~\cite{Koch:2008ia,Qiu:2011iv,Becattini:2021lfq}. 
The pertinent question then remains, whether we can pick out the smaller correlations 
due to a possible phase transition and eventually the QCD critical point~\cite{Braun-Munzinger:2015hba} after 
subtracting thermal emission and $v_2$, 
which are two ruling backgrounds~\cite{Heinz:2008tv,Koch:2008ia,Foka:2016zdb,Voloshin:2008dg,Snellings:2011sz,Rischke:2003mt}.
The perfect fluidity characteristics of QGP are studied in heavy-ion experiments~\cite{Shuryak:2003xe,Gyulassy:2004zy,Shuryak:2008eq},
with diverse experimental observables~\cite{Becattini:2021lfq}. This novel state of strongly
interacting matter produced in these high energy collisions show low shear
viscosity($\eta$) to entropy density($s$) ratio, $\eta/s$, which is in adjacent proximity to a nearly perfect
fluid~\cite{Heinz:2008tv,Becattini:2021lfq,Lacey:2006bc,Lacey:2009xx,Bagoly:2015ywa,Jia:2010zza,Lacey:2010ej}.

The paper is organised to start with a brief introduction of QGP produced 
from relativistic heavy-ion collisions, where correlations as well as fluctuations 
play an important role towards its eventual characterization.  The experimental results 
have already established that QGP is not a weakly coupled gas 
but a strongly coupled liquid~\cite{Heinz:2008tv,Becattini:2021lfq,Lacey:2006bc,Lacey:2009xx,Bagoly:2015ywa,Jia:2010zza,Lacey:2010ej}.
In Section 2 we have a brief survey of the existing results on correlation methodologies and an 
overview of the various aspects, which show that making a final conclusion on the 
nature of small systems is very tough and challenging. We
cannot characterize the A+A systems (or nucleus-nucleus collisions) successfully,
unless we have a detailed understanding of small systems like proton-proton (p+p) and proton-nucleus (p+A) 
collisions~\cite{Das:2021nqw}. Experimenters working on heavy-ion collisions, have only two quantities which they 
can regulate : which two nuclei they collide and at what energies~\cite{Rischke:2003mt,Braun-Munzinger:2015hba}. 
The high precision information on the colliding energies are available to experimentalists. 
But just knowing the colliding nuclei will not suffice, as is not the same as knowing the colliding systems in depth. 
The impact parameter b (which is the transverse distance between the
center of masses of the two nuclei) is not a measurable quantity. Also the location and motion of the nucleons in the
nuclei, including that of the quarks and gluons in the nucleons, are non-measurable quantities.
They have to be interpreted, as best as possible or as needed, event-by-event , from the observed
outcome of the collisions.  Any physical quantity measured experimentally is however, subject to fluctuations. 
Such fluctuations in general, depend on the properties of the system which is being studied (like for e.g here on the
properties of the fireball created in a heavy-ion collision) and may contain critical information
about the system. The original motivation for event-by-event studies in 
relativistic heavy-ion collisions has been to find hints for distinct event classes~\cite{Csernai:2012zf,Karsch:2005ps,Koch:2001zn,Qiu:2011iv}.
In particular it was felt that one can visualize events which would carry the signature of the QGP. But 
the recent QGP-like behaviour in p+p and p+A collisions (small systems) have challenged our 
age old understanding of high energy heavy-ion physics.  Section 3 is devoted to the discussion of 
why fluctuation is also an important methodology towards the understanding of small systems.
Finally, we summarise by looking into the emerging scope for such studies that lie ahead with ongoing and future endeavours.


\section{\bf{Correlations}}
\label{sec:correlations}

Two-particle correlations in high-energy collisions infer upon the important information for understanding 
QCD and have been explored previously for a wide range of collision 
energies in p+p, p+A, and A+A collisions. Such measurements 
can help to decipher the underlying mechanism of particle production 
along with the possible collective effects, which arise from the high particle densities available in such collisions.
Research of two-particle angular correlations are studied using two-dimensional 
$\Delta\eta-\Delta\phi$ correlation functions, where $\Delta\phi$ is the gap in azimuthal angle $\phi$ between the two particles and 
$\Delta\eta$ is the gap in the pseudo-rapidity which is defined as $\eta$ = - ln(tan($\theta$/2)). 
In this case the polar angle $\theta$ is described relative to the counter-clockwise beam.
Small systems like p+p and p+A collisions are not expected to see any correlation features as 
we have always considered them to be baseline and control systems to compare and understand 
A+A collisions. Hence, we do not expect any QGP-like phenomena in small systems.

However, contrary to our expectations, the long-range azimuthal correlations for 2.0 $<$ $ |\Delta\eta|$ $<$ 4.8 studies done 
by CMS experiment at $\sqrt{s}$ = 7 TeV, showed the long range ridge like structure at the near side ($\Delta\phi \approx 0$)
in p+p collisions~\cite{CMS:2010ifv}. In the high multiplicity domain of N $\approx$ 90 or higher, such clear feature emerged amid large rapidity 
differences ($ |\Delta\eta|$ $>$ 2), where most prominence was in the intermediate transverse momentum range, 1 $<$ $p_{T}$  $<$ 3 GeV/c.
Similarly in the high-multiplicity domain of p+Pb collisions at $\sqrt{s_{\rm NN}}$=5.02 TeV, the azimuthal correlations 
studies for 2.0 $<$ $|\Delta\eta|$ $<$ 4.0 by CMS also showed a qualitatively comparable long-range structure at the nearside $\Delta\phi$ $\approx$ 0.
In the domain of the same observed particle multiplicity, the comprehensive strength of the correlations is much larger in p+Pb collisions~\cite{CMS:2012qk}.
As observed in ALICE and ATLAS experiments, the long-range,  near-side angular correlations in particle production appeared in p+p~\cite{CMS:2010ifv} and also in p+Pb collisions~\cite{CMS:2012qk}. 
In p+Pb collision system, an away-side structure, located at $\Delta\phi$ $\approx$ $\pi$ 
and exceeding the away-side jet contribution, was noticed as well~\cite{Abelev:2012ola,Aad:2012gla} .
In the same p+Pb collision system at 5.02 TeV in another analysis, the two-particle angular correlations are also measured 
and expressed as associated yields per trigger particle by ALICE experiment~\cite{ALICE:2014mas}. For this analysis, the long-range pseudo-rapidity correlations 
have been subtracted from the per-trigger yields, to enable the understanding of the jet-like correlation peaks. The results show 
that the near-side and away-side jet-like yields are almost constant over a large range in multiplicity, other than the events with low multiplicity. 
Here we see a different feature, where the high multiplicity hard processes and number of 
soft particles have the same progression with multiplicity~\cite{ALICE:2014mas}. 

Prior to the LHC searches, ridge was observed at RHIC energies in high multiplicity (central) events in Cu+Cu collisions 
at $\sqrt{s}$ = 62.4 GeV~\cite{STAR:2011pct} and in Au+Au collisions at $\sqrt{s_{\rm NN}}$ = 200 GeV~\cite{Abelev:2009af}.
Such structure seen in heavy-ion collisions is also felt to arise due to collective flow~\cite{Luzum:2010sp}. 
The ridge has an azimuthal structure is long-range in rapidity, expanding over at least 5 even up to 10 units.  
Also the correlation is found to be collective and including almost all the particles produced in the event. 
A ridge correlation is not anticipated in a p+p collisions as the system is too dilute to create a fluid-like state.
This encouraged the researchers to pursue detailed analysis of the presence of collective phenomena
in p+p collisions which was known for a long time to happen only in heavy-ion collisions~\cite{Abelev:2009af}.
Thus its presence in small systems indicate the possibility of collective effects being present in such collisions.

So are the p+p collisions in LHC regime different? The two-particle charge-dependent correlations in pp, p+Pb, and Pb+Pb collisions 
as a function of the pseudo-rapidity and azimuthal angle difference, $\Delta\eta$ and $\Delta\phi$ are studied by ALICE experiment~\cite{ALICE:2015nuz}. 
These correlations are analysed using the balance function that examines the charge creation time along with the growth of collectivity in the produced system. 
The studies include the dependences of the balance function on the event multiplicity, along with the trigger and associated particle 
transverse momentum ($p_T$) in p+p, p+Pb, and Pb+Pb collisions at $\sqrt{s_{\rm NN}}$ = 7, 5.02, and 2.76 TeV.
The balance function is narrowed down both in $\Delta\eta$  and $\Delta\phi$ 
directions, in all the three systems for events with higher multiplicity, for 0.2 $<$ $p_{T}$ $<$ 2.0 GeV/c. The experimental findings compare well with the models 
that either incorporate some collective behavior (e.g. AMPT) or different mechanisms that lead to effects that resemble collective 
behavior (e.g. PYTHIA 8 with color reconnection). For higher values of transverse momenta the balance function 
becomes even narrower but does not show any multiplicity dependence, thereby pointing that the perceived narrowing 
with increasing multiplicity at low $p_{T}$ is an attribute of the bulk particle production. Thus we cannot rule out collectivity in small systems.

The anisotropic flow coefficients $v_n$($p_T$) are observables sensitive to the bulk properties 
of the created matter since they encode information about the equation of state and the transport properties of the medium.
There are three major genesis for the flow coefficients in collisions of identical nuclei at relativistic energies. 
Firstly, the more or less almond shape of the overlap region of the nuclei provides the prominent contribution to the $v_{2}$(elliptic flow) in non-central collisions.
Secondly, the event-by-event fluctuations in the positions of nucleons or in other terms of the colliding degrees of freedom inside the nuclei during 
the time of the collision can lead to changes of geometry. Thus deviating from the smooth shape corresponding to the overlap of the ideal spheres. 
These initial-state fluctuations contribute to triangular flow, $v_3$, and a rapidity-even (and hence the mid-rapidity) 
addition to directed flow, $v_1$, along with the $v_{2}$ measured in the most central collisions. 
They also add up to the higher harmonics, like $v_{4}$, $v_{5}$, $v_{6}$ (for e.g) which are, however, also majorly dominated by a 
third phenomenon, specifically attributed to the mixing of sizeable lower harmonics~\cite{Qiu:2011iv,Voloshin:2008dg,Snellings:2011sz}.


\section{\bf{Fluctuations}}
\label{sec:Fluctuations}

Event-by-event fluctuations and the pressure-driven
QGP evolution are encoded in the correlation measurements of
the flow magnitude $v_n$ and their phases $\Psi_{n}$. 
A Fourier expansion of the invariant triple differential
distributions are used for studying the patterns of anisotropic flow, 
which can be defined as:

\begin{equation}
\label{eq-flow}
E_{\textbf{p}}\frac{\dd^3N}{\dd^3\textbf{p}} = 
\frac{1}{2\pi}\frac{\dd^2 N}{p_T\,\dd p_T\,\dd y}\bigg[1 + \sum_n2v_n(p_T,y)\cos n(\varphi-\Psi_n)\bigg]
\end{equation}

where E is the energy of the particle, p the momentum, $p_{T}$ the transverse momentum,
$\varphi$ the azimuthal angle, y the rapidity, and $\Psi$ the reaction plane angle. The sine terms
in such an expansion go away due to the reflection symmetry with respect to the reaction
plane. The nth order flow (vector) $V_{n}$ is defined as: $V_{n}$ $\equiv$ $v_{n}$$e^{in\Psi_{n}}$~\cite{Qiu:2011iv,Voloshin:2008dg,Snellings:2011sz}.
The flow cumulants are always calculated for events with similar activity, in experiments.
Within the domain of a given activity measurement, fluctuations in the particle production process cause irreducible
centrality fluctuations, also known as volume fluctuations. Also as $v_{n}$ changes with centrality, centrality
fluctuations lead to an additional fluctuation of $v_{n}$, and thereby leading to a change in the flow cumulants.
Contemporary phenomenological studies of flow observables, have been a precision tool 
to elucidate the event-by-event quantum fluctuations in heavy-ion collisions.

Exploration for $p_{T}$-dependent flow vector fluctuations were performed by ALICE experiment 
by measuring $v_n$$\big\{$2$\big\}$$(p^a_T)$/$v_n$$[$2$]$$(p^a_T)$  and $r_{n}$(factorization factor)~\cite{Heinz:2013bua} in
Pb+Pb collisions at $\sqrt{s}_{\rm NN}$= 2.76 TeV and p+Pb collisions at $\sqrt{s}_{\rm NN}$= 5.02 TeV~\cite{ALICE:2017lyf}. 
The hydrodynamic calculations 
indicate a $p_{T}$ dependence of the flow vector $V_{n}$ due to event-by-event fluctuations in the initial energy density of the nuclear collisions.
Also the $p_{T}$-dependent flow vector fluctuations become crucial as they can unravel the initial conditions 
in heavy-ion collisions and hydrodynamic features in small systems like p+Pb where we have 
already observed collectivity in CMS, ALICE, ATLAS~\cite{CMS:2012qk,Abelev:2012ola,Aad:2012gla} and in the forward region by LHCb~\cite{LHCb:2015coe}.
In Pb+Pb collisions, both $v_n$$\big\{$2$\big\}$$(p^a_T)$/$v_n$$[$2$]$$(p^a_T)$  and $r_{2}$ show deviations from unity which reveal 
that these effects are more visible in the most central collisions and cannot be explained solely by non-flow effects. 
Hence, these results indicate the presence of possible $V_{2}$ vector fluctuations in Pb+Pb collisions~\cite{ALICE:2017lyf}. 
With relatively large statistical fluctuations in p+Pb collisions, deviations 
of $v_n$$\big\{$2$\big\}$$(p^a_T)$/$v_n$$[$2$]$$(p^a_T)$  and $r_{2}$ from unity are observed, as also seen in Pb+Pb system.
While comparing the measurements, done by ALICE experiment~\cite{ALICE:2019zfl}, of anisotropic 
flow coefficients $v_{n}$$\big\{${\it{k}}$\big\}$ of order $n$, obtained from ${\it{k}}$-particle correlations, and 
symmetric cumulants SC($m,n$) as a function of the produced particle multiplicity in 
small (p+p at $\sqrt{s}$ = 13 TeV and p+Pb collisions at $\sqrt{s_{\rm NN}}$ = 5.02 TeV) and large 
(Xe-Xe at $\sqrt{s_{\rm NN}}$= 5.5 TeV and Pb+Pb at $\sqrt{s_{\rm NN}}$ = 5.02 TeV) collision systems, we 
observe an ordering of the coefficients $v_{2}$ $>$ $v_{3}$ $>$ $v_{4}$ in small systems (p+p and p+Pb), 
similar to that seen in large collision systems. But even though a weak $v_{2}$ multiplicity dependence is observed relative 
to nucleus-nucleus collisions in the same multiplicity range, the observed 
long-range multi-particle azimuthal correlations in high multiplicity for small systems (p+p and p+Pb collisions) 
can neither be described by PYTHIA 8 nor by IP-Glasma+MUSIC+UrQMD model calculations.~\cite{ALICE:2019zfl}

The extent of the transverse flow, in small collision systems, is anticipated to be delicately tuned to the 
size of the initial source in the hydrodynamic model. Particularly, in p+Pb collisions, in the compact source scenario, 
the smaller source sizes are believed to yield substantial transverse flow and smaller initial eccentricities. 
The modified PCC (Pearson Correlation Coefficient) $\rho$($v_n$$\big\{$2$\big\}$$^2$,$[p_{T} $]), 
which quantifies the correlation between the flow harmonics and the mean transverse momentum was theoretically explained in ref~\cite{Bozek:2016yoj} 
and first measured by the ATLAS experiment~\cite{ATLAS:2019pvn} at the LHC in 5.02 p+Pb and Pb+Pb collisions.
In p+Pb collisions, the value of $\rho$($v_n$$\big\{$2$\big\}$$^2$,$[p_{T} $]) is negative and approximately independent of $N_{ch}$.
The negative sign of the modified Pearson correlation coefficient (PCC)~\cite{ATLAS:2019pvn} seems 
to infer towards the compact source scenario, and points to the role of the initial conditions in p+A systems. Even 
the higher p+Pb energies like 8.16 TeV collisions, where $v_{2}$, $v_{3}$, and $v_{4}$ Fourier 
coefficients are obtained from long-range two-particle correlations in event multiplicity classes, 
are found to be agreeable with 5.02 TeV p+Pb data in CMS experiment~\cite{CMS:2017kcs}. 
Normalized correlation coefficients for $v_{2}$ and $v_{3}$ are found to be quantitatively comparable between p+Pb and Pb+Pb, 
while for $v_{2}$ and $v_{4}$ the results are larger in p+Pb than in Pb+Pb. 
The corresponding result in 13 TeV p+p collisions shows a similar trend at high multiplicity 
but the large statistical uncertainties, hinder towards a quantitative statement~\cite{CMS:2017kcs}. 

Event-by-event fluctuations of thermodynamic quantities obtained and studied in high energy heavy-ion collisions 
provide a reasonable framework for exploring the nature of the QGP phase transition in the laboratory. 
This may be displayed in dynamical event-by-event fluctuations of the mean transverse 
momentum ($\langle p_{T} \rangle$) of final-state charged particles.
The multiplicity-dependent analysis of the event-by-event $\langle p_{T} \rangle$ fluctuations of 
charged particles in p+p collisions at $\sqrt{s}$ = 0.9, 2.76 and 7 TeV, 
and Pb+Pb collisions at $\sqrt{s_{\rm NN}}$ = 2.76 TeV are done with the ALICE detector at the LHC~\cite{ALICE:2014gvd}.
In the peripheral Pb+Pb collisions ($\langle$$dN_{ch}/d\eta$ $\rangle$ $<$ 100), the dependence of 
$\sqrt{C_m}/M(p_{\rm T})_m$ on $\langle$$dN_{ch}/d\eta$ $\rangle$ is very close to that 
observed in the p+p collisions at the corresponding collision energy. 
At larger multiplicities, the Pb+Pb data differ a lot from an extrapolation of 
the p+p results and we observe a strong decrease for $\langle$$dN_{ch}/d\eta$ $\rangle$ $>$ 500.
The dimensionless ratio $\sqrt{C_m}/M(p_{\rm T})_m$ quantifies the strength of the dynamical 
fluctuations in units of the average transverse momentum 
$M(p_{\rm T})_m$ in the multiplicity class $m$. The study in p+p collisions show 
characteristic decrease of  $\sqrt{C_m}/M(p_{\rm T})_m$ following a power law, where 
such a decrease is smaller than expected from a superposition of the independent sources and 
indicate towards multi-parton interactions (MPIs)~\cite{Diehl:2011yj}. Model comparisons of p+p 
data with PYTHIA show there there is no strong sensitivity of transverse momentum fluctuations 
to the processes of color reconnection. However, heavy-ion model like HIJING fail to explain the Pb+Pb data.


\section{Summary and Outlook}

The studies of correlations and fluctuations in small systems (p+p and p+A collisions)
can have interesting implications. The collective features observed in such collisions has 
convoluted the idea that a QGP may be formed in smaller collision systems.
Disagreement prevails as to whether such results are related to
QGP formation in small systems or whether they in fact disprove the existence of a thermalized medium and 
thereby hint that the experimentally observed flow signals in p+p, p+A  and A+A collisions 
can have the usual hydrodynamic, origin~\cite{Weller:2017tsr}. But that can be also 
contradicted, since the hydrodynamic simulations are yet not able to describe the multi-particle single and mixed harmonics cumulants~\cite{Zhou:2020pai}.
Also, many other features like jet quenching~\cite{Qin:2015srf}, bottomonium suppression~\cite{Das:2018xel} etc, are yet to be seen in small systems, 
including the suppression of high $p_{T}$ inclusive hadrons in p+A collisions with
respect to p+p collisions, which is also yet not measured. These results were first used to show the final-state QGP interactions
which caused the high $p_{T}$ hadron suppression in A+A collisions~\cite{Braun-Munzinger:2015hba}. 
The p+p and p+A collisions, with the increasing multiplicity, are now in a domain where the
macroscopic description (thermodynamics and hydrodynamics) can be applied.
The hydrodynamic models, when applied to p+A data, can explain many of the observed
features, but there are important counter arguments regarding the suitability of such applications~\cite{Shuryak:2013ke}.
Thus, a very elaborate description of a wide range of signatures, in an even comprehensive range of
colliding systems, will be very much needed to finally conclude towards a full understanding of these new challenging results.

The LHC has again started from 2022 and the data which 
will be collected in Run-3, will be an important addition for such studies on small systems~\cite{Das:2021bif}.
The experimental challenges of unmatched p+p luminosity, will be handled by the LHC experiments, with 
detector upgrades done during second long shutdown (LS-2). Equipped with such improved ability, the experiments can now 
isolate and precisely measure the products of the most interesting set off such high luminosity collisions. 
A more conclusive understanding will be also available with the results from p+Pb collisions.
As planned in Run-3 and 4, the high-multiplicity p+p collisions 
will be delivered by LHC~\cite{Noferini:2018are}. The LHC achieved nearly 30 $fb^{-1}$ by the end of 2012 and
plan to achieve 300 $fb^{-1}$ in its first 13-15 years of its operation.
The upgrading of the LHC injectors were accomplished 
during LS-2 before Run-3 and the next long shutdown 3 (LS-3) is currently planned 
to start in 2026 which may last for 3 years.
After LS-3, the LHC machine will move towards the High Luminosity (HL) configuration.
The High Luminosity LHC(HL-LHC) is a paramount
and arduous, upgrade~\cite{Apollinari:2017cqg}.
The  large p+p collision  data  sets  which are expected  to  be stored at
the  HL-LHC  will provide a fascinating setting
for these intriguing explorations~\cite{Noferini:2018are,Chapon:2020heu}.
Such higher multiplicity domains will enable to cover the difference between
the p+p and heavy-ion collisions, with new and fascinating detector
upgrades in LHC experiments~\cite{Citron:2018lsq}.


\bibliography{apssamp}

\end{document}